\documentclass[aps,prb,reprint,longbibliography]{revtex4-2}

\usepackage{graphicx}
\usepackage{siunitx}
\usepackage{bbm}
\usepackage{amsmath,amsfonts,amssymb,braket,bbold,mathtools,array}
\usepackage[colorlinks,citecolor=blue,linkcolor=blue,urlcolor=blue,bookmarks=false,hypertexnames=true]{hyperref}

\newcommand{\eps}{\varepsilon}
\renewcommand{\vec}[1]{\boldsymbol{#1}}
\newcommand{\ain}{a_\mathrm{in}}
\newcommand{\aout}{a_\mathrm{out}}
\newcommand{\taout}{\tilde a_\mathrm{out}}

\begin{document}

\title{Theory of the Simultaneous Transient Dispersive Readout of Multiple Spin Qubits}

\author{Florian Ginzel}
\author{Guido Burkard}
\affiliation{Department of Physics, University of Konstanz, D-78457 Konstanz, Germany}


\begin{abstract}
We propose a paradigm of multiplexed dispersive qubit measurement performed while the qubits dephase. A Laplace transformation of the time-dependent cavity response allows to separate contributions from multiple qubits coupled to the same resonator mode, thus allowing for simultaneous single-shot read out. With realistic parameters for silicon spin qubits we find a competitive readout fidelity, while the measurement time compares favourably to conventional dispersive readout. We extend the multiplexed readout method to quantum non-demolition measurements using auxiliary qubits.
\end{abstract}

\maketitle

\section{Introduction}

Fast high-fidelity readout is a key requirement to any qubit implementation~\cite{diVincenzoCriteria}, in particular in view of quantum error correction~\cite{RevModPhys.87.307,PhysRevLett.128.110504,Sundaresan2022}. The use of dispersively coupled microwave resonators~\cite{PhysRevA.69.062320} was pioneered by superconductiong qubits~\cite{PhysRevLett.95.060501,PhysRevLett.94.123602,PhysRevA.76.042319}. Dispersive readout makes use of a qubit-state dependent shift in the resonance frequency of the resonator. Similar techniques are now in reach for spin qubits in semiconductor quantum dots (QDs)~\cite{RMP-spinQubits,RevModPhys.79.1217} which have demonstrated strong spin-photon coupling~\cite{StrongCouplingPrinceton,Samkharadze1123,Landig2018} mediated by artificial spin-orbit coupling in a double QD (DQD)~\cite{PhysRevB.86.035314,PhysRevB.96.235434}. The dispersive readout of a single spin qubit was experimentally demonstrated~\cite{StrongCouplingPrinceton,PhysRevApplied.15.044052} and theoretically optimized~\cite{PhysRevB.100.245427,PhysRevB.99.245306}.

To scale up quantum processors and to facilitate fault-tolerant quantum computation, it is desirable to speed up qubit measurements by reading out multiple qubits simultaneously. With several qubits dispersively coupled to the same resonator, however, it is challenging to distinguish the contributions of the individual qubits~\cite{Nature449.443,PhysRevLett.102.200402,DiCarlo2010}, although parametric or dissipative dynamics allow a certain enhancement~\cite{PhysRevApplied.18.024009,Noh2021,PhysRevA.96.052321}. Readout with specialized electronics for each qubit~\cite{Neeley2010} introduces bulky components and thus limits scalability. In the context of superconducting qubits it is now common to use one readout resonator with individual frequency per qubit coupled to a shared microwave feedline~\cite{Jerger_2011,doi:10.1063/1.4739454,doi:10.1063/1.4764940,PhysRevLett.112.190504,PhysRevA.90.062333}. This approach is clearly not optimal for spin qubits, given their small size and their high density, compared to the space requirement of an on-chip microwave resonator. While gate-dispersive sensing~\cite{PhysRevLett.110.046805,PhysRevX.8.041032,West2019,Urdampilleta2019,Zheng2019,Crippa2019} has motivated research efforts towards multiplexed spin qubit readout~\cite{doi:10.1063/1.4868107,PRXQuantum.2.040306,Ruffino2022,Elhomsy2023}, a satisfying and reliable solution for fast simultaneous multi-spin readout is still lacking at the moment.

\begin{figure}[h]
\begin{center}
\includegraphics[width=0.5\textwidth]{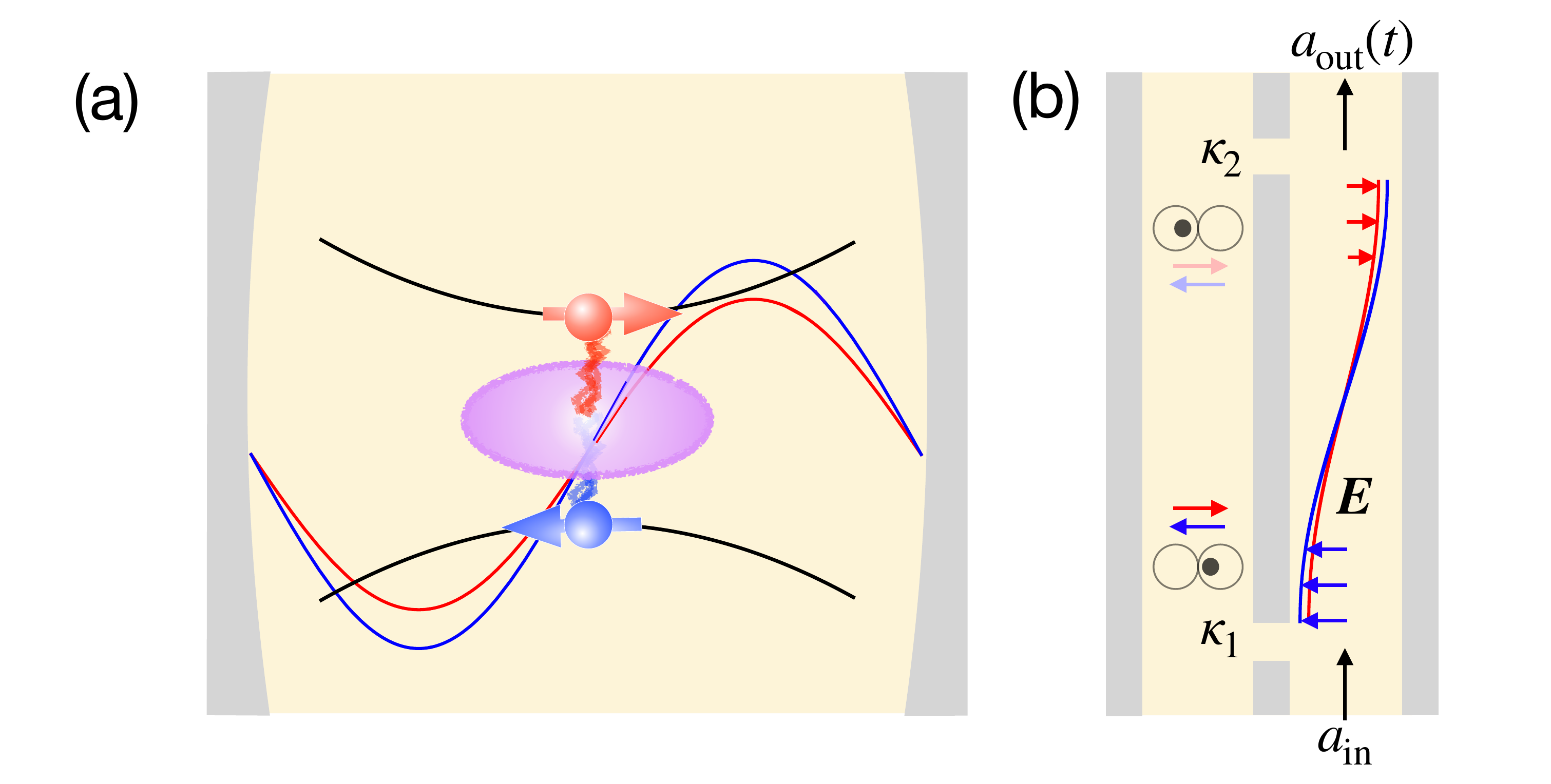}
\caption{(a) To measure the qubit (purple ellipse) it is off-resonantly coupled to a co-planar microwave cavity in the transient period while it dephases. The output field depends on the projection of the initial qubit state on the $x$-basis (red and blue arrows pointing left and right).
(b) Two qubits coupled to the cavity electric field $E$ are probed with the input field $\ain$. The output field $\aout$ depends on the initial state of both qubits and can analysed using heterodyne detection. The leakage rates $\kappa_{1(2)}$ represent the coupling of the resonator to the environment at port 1(2).\label{fig_systemsketch}}
\end{center}
\end{figure}

In this article we propose a cavity-based multi-qubit readout in the dispersive regime relying on transient qubit-cavity interaction terms, illustrated in Fig.~\ref{fig_systemsketch}. The signals stemming from different qubits coupled to the same resonator can be distinguished by a Laplace transform, allowing to read out multiple qubits simultaneously. We analyze the readout fidelity and discuss optimal operating regimes. The derivation of the underlying equations also applies to other qubit implementations whose Hamiltonian can be cast into the form of the dispersive Hamiltonian, Eq.~(\ref{eq_Hdispersive}), for example transmon qubits, however, we focus on silicon-based spin qubits, since we estimate that this platform can benefit the greatest for our results.

This article is organized as follows. In Sec.~\ref{sec_model} the model of an ensemble of spin qubits dispersively coupled to a single resonator mode is introduced. In Sec.~\ref{sec_procedure} the proposed measurement procedure is outlined and, subsequently, in Sec.~\ref{sec_fidelity} the readout fidelity is derived for a single-qubit readout and then generalized to the case of multiple qubits.

\section{Model\label{sec_model}}

We consider multiple DQDs in an inhomogeneous magnetic field, where the $\mu$th  DQD is occupied by a single electron and is described by $H_d^{(\mu)} = H_m^{(\mu)} + H_z^{(\mu)}$, with 
\begin{eqnarray}
H_m^{(\mu)} &=& \frac{1}{2}\eps_\mu \tau_z'^{(\mu)} + t_c^{(\mu)} \tau_x'^{(\mu)},\\
H_z^{(\mu)} &=& \frac{1}{2} B_z^{(\mu)}\sigma_z'^{(\mu)} + \frac{1}{2} b_x^{(\mu)} \tau_z'^{(\mu)}\sigma_x'^{(\mu)},
\end{eqnarray}
where $\vec{\tau'}^{(\mu)}$ ($\vec{\sigma'}^{(\mu)}$) is the vector of Pauli operators for the position (spin) of the electron. Then, $\eps_\mu$ is the energy detuning between the left and right QD levels, $t_c$ is the spin-conserving tunneling matrix element, $B_z^{(\mu)}$ is the Zeeman splitting, and $b_x^{(\mu)}$ is the difference in transverse magnetic field between the QDs in energy units~\cite{PhysRevB.96.235434,PhysRevB.100.245427}.

We assume that all DQDs couple to a single cavity mode, $H_c = \omega_0 a^\dagger a$,  via the electric dipole interaction 
\begin{equation}
H_i^{(\mu)} = g_c^{(\mu)} \tau_z'^{(\mu)} (a^\dagger + a),
\end{equation}
where $a^{(\dagger)}$ annihilates (creates) a photon, $\omega_0$ is the resonator frequency, and $g_c^{(\mu)}$ is the charge-photon coupling strength \cite{Cohen-Tannoudji,ScullyZubairy}. The leakage rates at the resonator ports $i=1,2$ are defined as $\kappa_i$ and the total leakage rate is $\kappa = \kappa_1+\kappa_2$. Furthermore, a probe field,
\begin{equation}
H_\mathrm{in} = i\sqrt{\kappa_1}\left( \ain e^{-i\omega_p t}a^\dagger - \ain^* e^{i\omega_p t}a  \right),
\end{equation}
with frequency $\omega_p$ and amplitude $\ain$ is injected into port 1 of the resonator.

Transforming the system $H=H_c + \sum_\mu (H_d^{(\mu)}+H_i^{(\mu)})$ into the eigenbases of the DQDs, $H_d^{(\mu)}$, \cite{PhysRevB.100.245427}  allows defining a spin qubit with Pauli operators $\vec\sigma^{(\mu)}$ in the orbital ground state of each DQD while the synthetic spin-orbit coupling $b_x$ gives rise to the indirect spin-photon coupling $g_s^{(\mu)} \approx g_c^{(\mu)} (2t_c^{(\mu)})^2 b_x^{(\mu)}/\Omega_\mu [\Omega_\mu^2 - (B_z^{(\mu)})^2]$~\cite{PhysRevB.96.235434,PhysRevB.100.245427}. Here $\Omega_\mu$ denotes the orbital energy splitting $\sqrt{\eps_\mu^2 + 4(t_c^{(\mu)})^2}$ of qubit $\mu$. Note that an alternative coupling mechanism has been proposed that goes beyond the linear response of the qubit, giving rise to a longitudinal coupling which depends on the resonator frequency, and which can also mediate a dispersive-like qubit readout~\cite{PhysRevB.99.245306}. The DQD with a single charge has a strong electric dipole moment, however, such that the transverse coupling in $H_i^{(\mu)}$ is by far the dominating effect. Neglecting additional smaller contributions is supported by experimental observations on comparable devices which were described by the coupling discussed here to a high accuracy~\cite{StrongCouplingPrinceton,PhysRevApplied.15.044052}.

To describe cavity-based readout of the spin qubits we perform a rotating wave approximation and apply a Schrieffer--Wolff transformation~\cite{SW,SW_math} to model the dispersive regime where all electronic transitions are off-resonant from the probe field~\cite{PhysRevA.69.062320}. In the spin-like subspace we find ~\cite{PhysRevB.100.245427},
\begin{equation}
H \approx \delta_c a^\dagger a + \sum_\mu [\delta_s^{(\mu)}/2 - \chi_s^{(\mu)}(a^\dagger a + 1/2)]\sigma_z^{(\mu)}.\label{eq_Hdispersive}
\end{equation}
Here, $\delta_c=\omega_0-\omega_p$ [$\delta_s^{(\mu)} = E_s^{(\mu)} -\omega_p$] is the detuning of the resonator frequency (spin qubit splitting $E_s^{(\mu)}$) from the probe field,  $\chi_s^{(\mu)} \approx (g_s^{(\mu)})^2/\Delta_\mu$ is the dispersive shift due to the spin $\mu$ with $\Delta_\mu = \omega_0-E_s^{(\mu)}$. Note that the Hamiltonian in Eq.~(\ref{eq_Hdispersive}) is not unique to spin qubits and could also describe, for example, superconducting transmon qubits~\cite{PhysRevA.69.062320}. 

The equations of motion for $a$ and $\sigma_-^{(\mu)} = \frac{1}{2}(\sigma_x^{(\mu)} - i \sigma_y^{(\mu)})$ are finally obtained by including the incoherent interactions derived from input-output theory~\cite{PhysRevB.100.245427}
\begin{eqnarray}
    \dot a &\approx & s_c a - \sqrt \kappa_1 \ain - \sum_\mu \frac{\kappa g_s^{(\mu)}}{2\Delta_\mu} \sigma_-^{(\mu)}, \label{eq_adot}\\
    \dot \sigma_-^{(\mu)} &\approx & s_\mu  \sigma_-^{(\mu)} + \frac{g_s^{(\mu)}}{\Delta_\mu} \sqrt{\kappa_1} \ain \sigma_z^{(\mu)}, \label{eq_sigma-dot}
\end{eqnarray}
with $s_c = - i \left(\delta_c - \sum_\mu \chi_s^{(\mu)} \sigma_{z,0}^{(\mu)}\right) - \kappa/2$ and $s_\mu = -i \left[ \delta_s^{(\mu)} - 2 \left( a^\dagger a + \frac{1}{2}\right) \chi_s^{(\mu)} \right] - \gamma_\mu /2$, and where $\gamma_\mu$ is the dephasing rate of qubit $\mu$ and $\sigma_{i,0}^{(\mu)}=\sigma_{i}^{(\mu)}(t=0)$. The output field at port 2 of the cavity is derived from input-output theory~\cite{PhysRevA.30.1386,PhysRevA.31.3761,PhysRevB.100.245427},
\begin{equation}
    \aout \approx \sqrt{\kappa_2} a + \sum_\mu \sqrt{\kappa_2} \frac{g_s^{(\mu)}}{\Delta_\mu} \sigma_-^{(\mu)}. \label{eq_aout}
\end{equation}

The last terms in Eqs.~(\ref{eq_adot}) and (\ref{eq_aout}) are usually neglected in the treatment of dispersive readout since they vanish in the stationary state where the qubit is dephased and in a typical measurement setup they are removed by frequency filters to reduce noise. However, the dependence of the output field $\aout(t)$ on $\sigma_-^{(\mu)}$ before the steady state is reached implements a measurement in the $x$-basis of qubit $\mu$ if the detection is sufficiently broadband. This readout must be considered destructive as the qubit dephases in the process. It is possible to turn this readout into a quantum non-demolition (QND) measurement in the $z$-basis by introducing an ancilla qubit. The protocol is in analogy to the recently demonstrated QND measurement in Ref.~\cite{Nakajima2019}. The ancilla $a$ is initialized in state $|0_z\rangle_a$. A CNOT gate with the qubit as control and the ancilla as target, followed by a Hadamard gate on the ancilla prepares the ancilla in the $x$-basis, conditional on the state of the qubit in the $z$-basis. The qubit is then quickly detuned to a protected idling spot~\cite{PhysRevB.100.125430,Fehse2022} while the ancilla is coupled to the resonator and read out using the transient dispersive readout. Due to the entangling gate, the state of the qubit can be inferred from this measurement, although an imperfect gate may introduce uncertainty.  

Contributions from multiple qubits can be separated in frequency~\cite{SignalProcessingBook} and thus observed simultaneously by detuning the qubit splittings $E_s^{(\mu)}$ from each other. This detuning will also suppress coherent oscillations between different qubits mediated by the resonator~\cite{PhysRevA.69.062320}, justifying their negligence in Eqs.~(\ref{eq_adot}) and (\ref{eq_sigma-dot}).

\section{Proposed readout procedure\label{sec_procedure}}

Here, we investigate the expectation values of the system observables as provided by the input-output theory described in the previous section. We note that the dispersive qubit-photon interaction in Eq.~(\ref{eq_Hdispersive}) commutes with the resonator Hamiltonian $H_c$ and no absorption or emission of photons occurs. Furthermore, we assume that the cavity and the qubits are initially decoupled, which can be realized by detuning all qubits far from the charge transition, $|\eps_\mu /2 t_c^{(\mu)}| \gg 1 $ and it allows switching on or off the probe field without disturbing the qubits. After the ring-up of the cavity, the qubits can be pulsed to the readout configuration. Thus, we can assume that the photon number of the resonator is initialized to its steady state value and that it can be treated as a constant.

Relying on the assumptions that the photon number is a constant of motion during the readout, $(a^\dagger a) (t) \approx (a^\dagger a) (0)$, and that there is no initial spin-photon correlation, $\langle (a^\dagger a) (0) \sigma_-^{(\mu)}(0)\rangle = \langle (a^\dagger a) (0)\rangle \langle \sigma_-^{(\mu)}(0)\rangle$, because the interaction between DQD and resonator is switched on at time $t=0$ the Laplace transform~\cite{DoetschLaplace} of the expectation value of the output field,
\begin{eqnarray}
    \taout(s)\!\! &\approx& \!\! \sqrt{\kappa_1} \tilde a (s) + \frac{\sqrt{\kappa_2}}{\kappa} \sum_\mu \eta_\mu (s),\label{eq_taout}\\
    \eta_\mu(s)\!\! &=& \!\!\frac{\kappa g_s^{(\mu)} }{2\Delta_\mu}\! \left(\!\sigma_{-,0}^{(\mu)} -\! \sqrt{\kappa_1} \ain \frac{g_s^{(\mu)} \sigma_{z,0}^{(\mu)} }{\Delta_\mu s} \! \right)\!\!/(s - s_\mu),\label{eq_eta}
\end{eqnarray}
can be found from the equations of motion, Eqs.~(\ref{eq_adot}) and (\ref{eq_sigma-dot}). The term including $\sigma_{z,0}^{(\mu)}$ in $\eta_\mu$ can be considered irrelevant for the readout, since for typical spin qubit parameters in the dispersive regime it is expected to be roughly two orders of magnitude smaller than the leading term~\cite{StrongCouplingPrinceton}. The field $\tilde a$ as a function of the complex frequency $s$ is given by
\begin{equation}
    \tilde a (s) \approx \left(a_0 - \sqrt \kappa_1 \ain/s -\sum_\mu \eta_\mu (s) \right)/(s - s_c).
\end{equation}
The field operator at the initial time is given by $a(t=0)=a_0$. For the qubit $\mu=1,2...$, $\taout(s)$ exhibits a singularity at $s_\mu$. The slope near the singularity depends on the initial state $\sigma_{-(z),0}^{(\mu)}$. For the basis states of the $x$-basis this is $\sigma_{-,0}^{(\mu)} =\pm 1/2$ and $\sigma_{z,0}^{(\mu)} =0$.

\begin{figure*}[h]
\begin{center}
\includegraphics[width=0.95\textwidth]{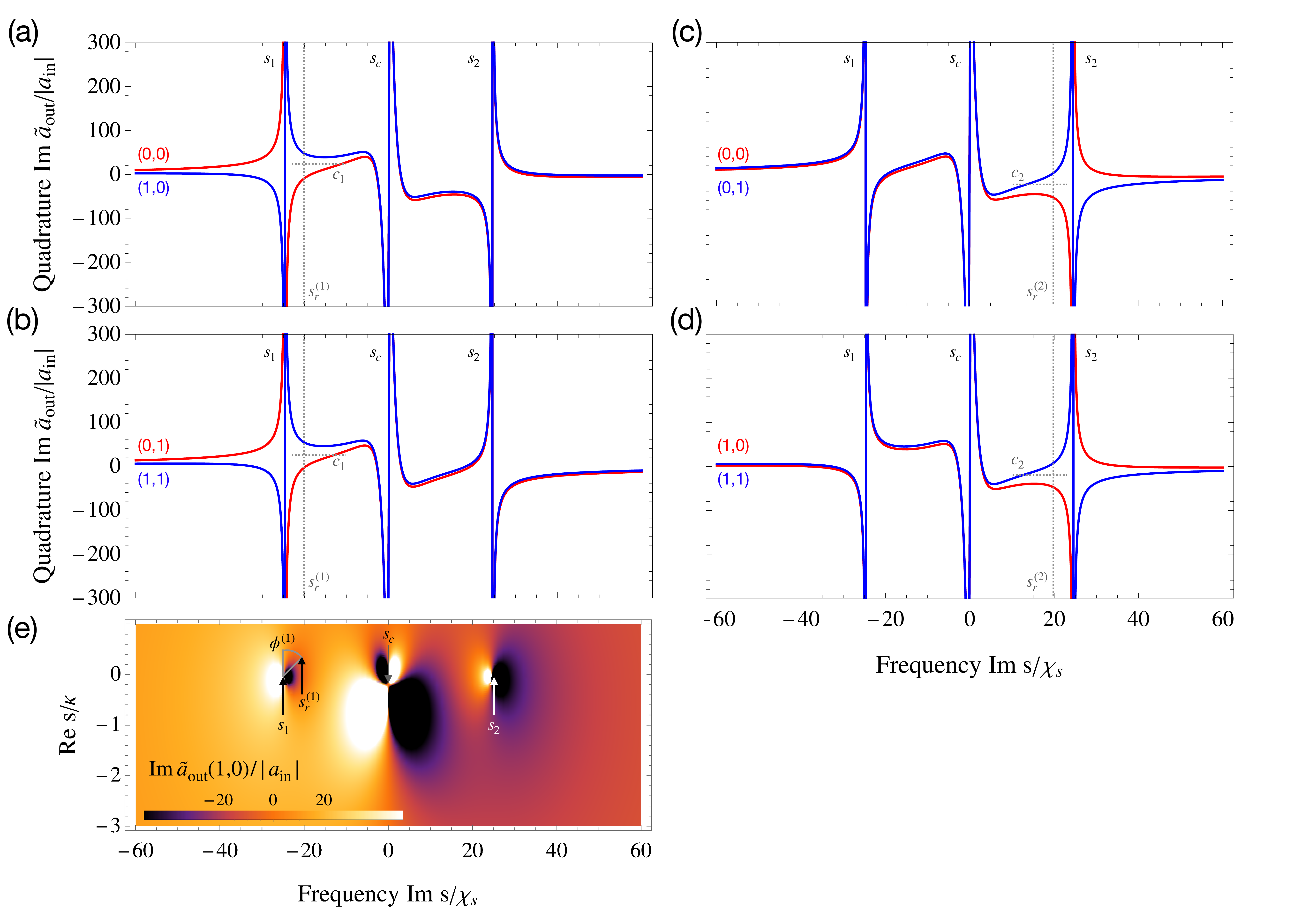}
\caption{Laplace transform of the quadrature $\mathrm{Im}\, \aout$ of the output field with two qubits at $2\mathrm{Re}\, s = \gamma_1 = \gamma_2$. (a) The result for the initial qubit states $\ket{0_x}_1 \ket{0_x}_2$ ($\ket{1_x}_1 \ket{0_x}_2$) and (b) for the initial states $\ket{0_x}_1 \ket{1_x}_2$ ($\ket{1_x}_1 \ket{1_x}_2$) is depicted in red (blue). For each qubit $\mu =1,2$ a pole $s_\mu$ of $\mathrm{Im}\, \taout$ is observed, the center pole $s_c$ is the cavity resonance. Directly at the poles $\mathrm{Im}\, \taout$ diverges and no distinction between the qubit states can be made. However, in the vicinity of $s_\mu$ the initial state of qubit $\mu$ can be revealed: First, $\mathrm{Im}\, \taout$ is evaluated at a readout frequency $s_r^{(\mu)}$ (vertical dashed gray line for qubit 1) near the pole. Then the obtained value is compared to a threshold $c_\mu$ (horizontal dashed gray line), the initial qubit state is assigned depending whether $\mathrm{Im}\, \taout(s_r^{(\mu)})$ is greater or less than $c_\mu$. (c), (d) Readout of qubit 2 depending on the initial state of qubit 1. In analogy to panels (a) and (b) a readout frequency $s_r^{(2)}$ and a threshold $c_2$ are chosen, which allow distinguishing the initial states of qubit 2 from the cavity response. As the figure shows, this is possible for both qubits independently.
(e) $\mathrm{Im}\, \taout$ in the complex plane for the initial state $\ket{0_x}_1\ket{0_x}_2$. The poles $s_{1(2)}$ and $s_c$ belonging to qubits 1 and 2 and the cavity are indicated, as well as the complex readout frequency $s_r^{(1)}$ for qubit 1.\label{fig_scheme}}
\end{center}
\end{figure*}

We assume that the quadrature $\mathrm{Im}\, \aout$ of the transmitted output field is measured \cite{PhysRevB.100.245427} and derive its Laplace transform with the aid of the identity $\mathcal L f^*(s) = (\mathcal L f)^*(s^*)$~\cite{DoetschLaplace}. An example for $\tilde a_{\rm out}(s)$ with two qubits is shown in Fig.~\ref{fig_scheme}. It is clearly possible to distinguish the initial state of each qubit $\mu=1,2$ by choosing a proper complex readout frequency $s_r^{(\mu)} = s_\mu + \Delta s_\mu$ in the complex plane and evaluating $\mathrm{Im}\, \taout (s_r^{(\mu)})$. The measurement outcome for the qubit state is assigned depending on whether the observed $\mathrm{Im}\, \taout (s_r^{(\mu)})$ is above or below a threshold $c_\mu$. The choice of $\Delta s_\mu$ affects the contrast between the possible measurement outcomes and therefore the fidelity, as will be discussed in Sec.~\ref{sec_fidelity}. In the multi-qubit case certain restrictions for $\Delta s_\mu$ arise, which are captured by Eq.~(\ref{eq_separation}).

All plots are drawn for realistic values of $\omega_0=\omega_p=\SI{23.6}{\micro\electronvolt}$, $B_z^{(1)}=\SI{23.6}{\micro\electronvolt}$, $B_z^{(2)}=\SI{23.664}{\micro\electronvolt}$, $g_c^{(\mu)}=\SI{0.16}{\micro\electronvolt}$, $b_x^{(\mu)}=\SI{1.68}{\micro\electronvolt}$, $t_c^{(\mu)}=\SI{20}{\micro\electronvolt}$, $\eps_\mu = 0$, $\kappa_i = \SI{7}{\nano\electronvolt}$ and $\gamma_\mu=\SI{1.65}{\nano\electronvolt}$~\cite{StrongCouplingPrinceton}. With this choice the qubit-qubit interaction $J_{12}$ mediated by the resonator can thus be safely neglected, since $J_{12} / |E_s^{(1)} - E_s^{(2)}| \approx 0.009 \ll 1$~\cite{Nature449.443}. The input power $|\ain|$ is chosen such that the steady state cavity population is no more than 5\% of the critical photon number for the dispersive approximation~\cite{PhysRevLett.105.100504,PhysRevB.100.245427}.

To separate the singularities $s_\mu$ of the individual qubits along the imaginary axis of the complex frequency space the qubit frequency $E_s^{(\mu)}$ can be tuned by means of $B_z^{(\mu)}$ and $t_c^{(\mu)}$ to set $\delta_s^{(\mu)}$ and $\chi_s^{(\mu)}$ for each qubit. The electrostatic detuning of the DQDs $\eps_\mu$ has a weaker effect since $\partial_{\eps_\mu} E_s^{(\mu)}|_{\eps_\mu = 0} = 0$ \cite{PhysRevB.96.235434,PhysRevB.100.125430,PhysRevB.100.245427}. By tuning the qubit away from its sweet spot $\eps_\mu = 0$, however, the dephasing $\gamma_\mu$ can be enhanced~ \cite{PhysRevB.100.125430}. This offers a possibility to separate the contributions $s_\mu$ along the real axis.

\begin{figure}
\begin{center}
\includegraphics[width=0.45\textwidth]{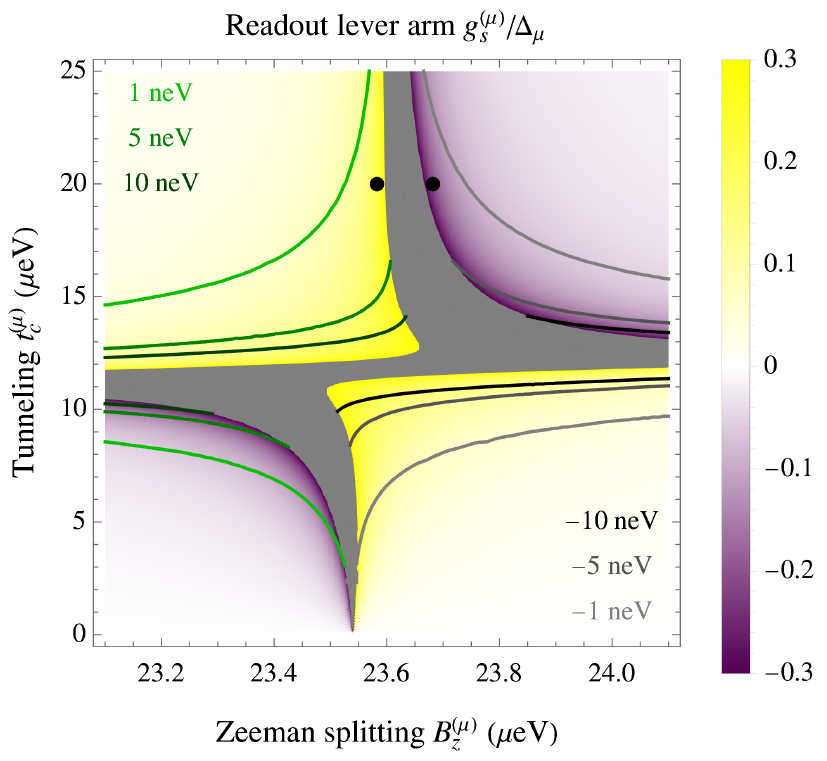}
\caption{Density plot of $g_s^{(\mu)}/\Delta_\mu$ which is the lever arm of $\eta_\mu$  for a single qubit. The contour lines indicate a constant dispersive shift $\chi_s^{(\mu)}$. For ideal multi-qubit readout the qubits have distinct $\chi_s^{(\mu)}$ and a high $|g_s^{(\mu)}/\Delta_\mu|$ similar for all $\mu$, so their singularities appear well separated with high visibility. Outside the gray shaded parameter range the dispersive approximation is valid, whereas within the gray region deviations become larger until ultimately the dispersive approximation breaks down when qubit and cavity become resonant. The black dots indicate the parameters for the two qubits in the example of Figs.~\ref{fig_scheme} and \ref{fig_fidelity}. This choice of parameters is motivated by the unambiguous observation of dispersive spin-photon interaction with the values of spin-cavity detuning at these spots~\cite{StrongCouplingPrinceton}.\label{fig_DQDparameters}}
\end{center}
\end{figure}

Tuning $B_z^{(\mu)}$ and $t_c^{(\mu)}$ to separate the qubits will also alter $g_s^{(\mu)}/\Delta_\mu$, which appears as a prefactor in $\eta_\mu$ Eq.~(\ref{eq_eta}) and which we denote as the readout lever arm. In Fig.~\ref{fig_DQDparameters} we plot $g_s^{(\mu)}/\Delta_\mu$ and contours of constant $\chi_s^{(\mu)}$, showing that it is indeed possible to choose different $\chi_s^{(\mu)}$ with comparable $|g_s^{(\mu)}/\Delta_\mu|$. Note, however, that $|g_s^{(\mu)}/\Delta_\mu| \ll 1$ is required to keep the spin-photon interaction off-resonant.

In an experiment, $\mathrm{Im}\, \taout (s)$ can be obtained by performing a series of short heterodyne measurements of $\mathrm{Im}\,\aout (t)$ with a duration of $t_i \ll  \min_\mu (\mathrm{Im}\,s_\mu)^{-1} , \min_\mu 2 \gamma_\mu^{-1}$, for a total readout time $t_r \gtrsim \max_\mu (\mathrm{Im}\,s_\mu)^{-1} , \max_\mu 2 \gamma_\mu^{-1}$. The numerical Laplace transform of the discrete time data can be implemented by the $z$-transformation and yields a continuous function in the complex frequency space~\cite{z-transform}. We assume that the choice of the readout frequency $s_r^{(\mu)}$ in the complex plane is limited only by the accuracy of the frequency measurement and thus $|\Delta s_\mu| = \kappa$.

\section{Readout fidelity\label{sec_fidelity}}

Equation~(\ref{eq_taout}) yields only the expectation value of the output field. Assuming that the fluctuations are Gaussian we also derive the variance
\begin{eqnarray}
    \sigma^2 (s) &\approx& 
    \langle {\rm Im}\, \taout(s)^2\rangle
    -\langle {\rm Im}\, \taout(s)\rangle ^2
    \nonumber\\
    & = &
    \left| \frac{1/R t_i + \frac{k_2}{2}\sum_\mu \xi_\mu}{s - 2 s_c} \right|, \\
    \xi_\mu &=& \left( \frac{\kappa g_s^\mu}{\Delta_\mu} \right)^2 \left[F_g^{(\mu)}-(F_g^{(\mu)})^2 \right] \left\{ s - 2 s_\mu
    \right\}^{-1} \nonumber\\
    && \times\left\{ s - 2 s_\mu + 2i \left[ \delta_c - \chi_s \sigma_{z,0}^\mu \right] + \frac{\kappa - \gamma^\mu}{2} \right\}^{-1},
\end{eqnarray}
of the measured quadrature in the complex frequency space. Here, we have used that a quadrature of the output field can be be measured with accuracy $\sigma^2(t) = 1/R t$ in a given time $t$, where $R = 4/(2 \bar N + 2 N_\mathrm{amp} + 1)$ with the number of thermal noise photons $\bar N$, and the number of noise photons added by the detector $N_\mathrm{amp}$~\cite{PhysRevB.100.245427}. We assume that the input field is limited by vacuum fluctuations ($\bar N = 0$) and that the detector is quantum limited ($N_\mathrm{amp} = 1/2$)~\cite{RevModPhys.82.1155} to estimate the optimal fidelity. The terms $\xi_\mu$ emerges when an ancilla is entangled with the qubit to allow a QND measurement, with a fidelity  $F_g < 1$ of the entangling two-qubit gate. The $\xi_\mu$ vanish for $F_g = 1$ or if no ancilla is used and can be understood as being due to the uncertainty of the state of the ancilla after an imperfect conditional rotation. In the framework of the Laplace transform this term represents the dynamics of $\mathrm{var}\,a$ due to a non-zero initial variance of the ancilla operators.

First, we estimate the readout fidelity for a single qubit, for simplicity. We define a threshold $c$ to discriminate $\mathrm{Im}\,\taout(s_r,\sigma)$, where $\sigma=0,1$ is the initial qubit state. The initial qubit state is identified as $\ket{0_x}$ [$\ket{1_x}$] upon observation of $\mathrm{Im}\,\taout(s_r,\sigma) > c$ with $\taout(s_r,1) < \taout(s_r,0)$ [$\taout(s_r,0) < \taout (s_r,1)$] and as $\ket{1_x}$ [$\ket{0_x}$] otherwise. The probability to incorrectly find $\mathrm{Im}\,\taout(s_r,\sigma) > c$ is given by $\alpha_0$ and the probability to incorrectly find $\mathrm{Im}\,\taout(s_r,\sigma) < c$ is $\alpha_1$, defined as
\begin{equation}
\alpha_\sigma = \frac{1}{2} \left[1 \pm (-1)^\sigma
\mathrm{erf}\left( \frac{\mathrm{Im}\,\taout(s_r,\sigma)- c}{\sqrt{2\sigma^2(s_r,\sigma)}} \right)\right].\label{eq_alpha}
\end{equation}
The upper sign should be used if $\taout(s_r,0) < \taout(s_r,1)$, the lower sign otherwise. The choice $c=\frac{1}{2}[ \taout(s_r,0)+ \taout(s_r,1) ]$ minimizes $\alpha_\sigma$~\cite{2004Natur.430..431E}. 

Equation~(\ref{eq_alpha}) is valid for a direct $x$-measurement of a single qubit. If an entangling gate with fidelity $F_g\leq 1$ is used to measure the qubit via an ancilla, then the readout fidelity $F_r\le F_g$, necessitating optimized gates~\cite{PhysRevLett.118.150502,PhysRevB.105.245308}. We account for this by introducing $\alpha_0' = F_g \alpha_0 + (1-F_g)(1-\alpha_1)$ and $\alpha_1' = F_g \alpha_1 + (1-F_g)(1-\alpha_0)$, where the first (second) term corresponds to readout   (two-qubit gate) error. We finally define the readout infidelity $1-F_r= \frac{1}{2}(\alpha_0' + \alpha_1')$~\cite{PhysRevB.74.195303,Nakajima2019}.

\begin{figure}
\begin{center}
\includegraphics[width=0.4\textwidth]{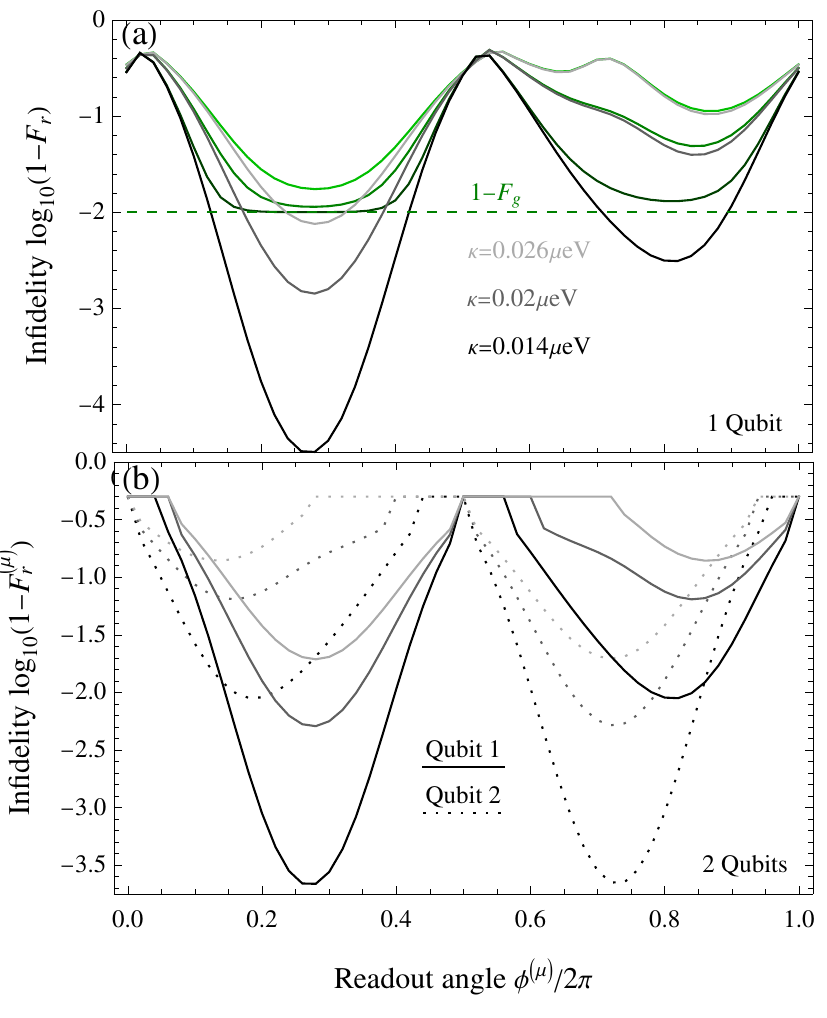}
\caption{(a) Single-shot readout fidelity $F_r^{(\mu)}$ as a function of $\phi^{(\mu)}=\arg \Delta s_\mu$ for one qubit. The shade of the line indicates the cavity output rate $\kappa$. The gray (green) lines correspond to a destructive direct measurement (indirect QND measurement with $F_g =0.99$). The asymmetry is due to the relative placement of $s_r^{(\mu)}$ and $s_c$. If $\kappa \gg \gamma_\mu$ the readout frequency $s_r^{(\mu)}$ and the singularity $s_c$ may approximately coincide, rendering readout at that angle impossible.
(b) Readout fidelity for simultaneous two-qubit readout. Solid (dotted) refers to qubit $\mu=1(2)$. The angles $\phi^{(1)}$ and $\phi^{(2)}$ are independent. There are choices for $\phi_\mu$ where qubit $\mu$ cannot be read out since the effect of the other qubit is too strong. This is particularly prominent at large $\kappa$, indicated by the lightest shade.
In both panels $t_r/t_i=4000$. The qubit parameters are taken from the experimental realization in Ref.~\cite{StrongCouplingPrinceton} and shown in Fig.~\ref{fig_DQDparameters}.\label{fig_fidelity}}
\end{center}
\end{figure}

In Fig.~\ref{fig_fidelity}(a) the single-qubit readout infidelity is plotted as a function of $\phi = \arg \Delta s$. The optimal readout fidelity is achieved if $\Delta s$ is imaginary, due to the shape of the singularity. The orientation of $s_r$ with respect to $s_c$ gives rise to an anisotropy in complex frequency space. For $\kappa \gg \gamma$ the readout becomes impossible if $s_r$ faces towards $s_c$ since in that case $s_r \approx s_c$, where no distinction between the qubits states is possible [see Fig.~\ref{fig_DQDparameters}(c)]. Generally, a smaller $\kappa$ grants lower $1-F_r$ because it reduces the frequency uncertainty and also increases $\mathrm{Im}\,\taout(s_r,0)/\sqrt{2\sigma^2(s_r,0)}$. Furthermore, $\Delta = \omega_r - E_s$ should be as small as agreeable with the dispersive approximation to boost $F_r$.

We now turn back to the multi-qubit case. To generalize the fidelity to multiple qubits it must be taken into account that $\alpha_\sigma^{(\mu)}$ is conditional on the state of the other qubits. To estimate the single-qubit readout fidelity $F_r^{(\mu)}$ of qubit $\mu$ in a simultaneous measurement of two qubits, Fig.~\ref{fig_fidelity}(b), we average the threshold $c_\mu$ over all states of the other qubit. In our example the two qubits are symmetrically detuned from the cavity frequency and thus their singularities are mirrored copies of each other.

Remarkably, there are choices of $\phi_\mu$ where qubit $\mu$ cannot be read out with our scheme. For example, if $\mathrm{Im}\, \taout(s_r^{(2)},0,0)< \mathrm{Im}\, \taout(s_r^{(2)},0,1)<\mathrm{Im}\, \taout(s_r^{(2)},1,0)<\mathrm{Im}\, \taout(s_r^{(2)},1,1)$ it is impossible to discern the basis states of qubit 2 with a single unconditional threshold $c_2$. To identify choices of $s_r^{(2)}$ where this problem does not occur we demand $\zeta_2 [ \mathrm{Im}\, \taout ( s_r^{(2)},0,1 ) - \mathrm{Im}\, \taout ( s_r^{(2)},1,0 ) ] > \bar\sigma (s_r^{(2)})$ where $\zeta_2 = \mathrm{sgn} [ \mathrm{Im} \, \taout ( s_r^{(2)},1,1 ) - \mathrm{Im}\, \taout ( s_r^{(2)},0,0 )]$ is a sign dependent on the DQD parameters and $\bar\sigma$ is $\sigma$ averaged over all two-qubit states. Using also the analogous condition for qubit 1 we find that both qubits can be read out if for $\mu=1,2$
\begin{equation}
    \zeta_\mu \frac{\sqrt{\kappa_2}}{\kappa} \, \mathrm{Im} \Bigg[ \frac{s_r^{(\mu)} + i \delta_c }{s_r^{(\mu)} - s_c}  \left( \frac{1}{\Delta s_\mu} - \frac{1}{s_r^{(\mu)} - s_{\bar\mu}} \right) \Bigg] > \frac{\bar \sigma (s_r^{(\mu)}) }{2}.\label{eq_separation}
\end{equation}
This provides a rule how far the readout frequency $s_r^{(\mu)}$ of qubit $\mu$ must be from the singularity $s_{\bar\mu}$ of the other qubit and from $s_c$. Here, $\bar\mu=2 (1)$ if $\mu=1(2)$.

Note that Eq.~(\ref{eq_separation}) is valid for the two-qubit example of Fig.~\ref{fig_fidelity}(b) but not the general case. To include more qubits, the term $1/(s_r^{(\mu)} - s_{\bar\mu})$ should be replaced by the sum $\sum_{\nu\neq\mu} 1/(s_r^{(\mu)} - s_{\nu})$. Together with the tuneability of $E_s^{(\mu)}$ and $\gamma_\mu$ in a given device this complex frequency qubit spacing determines how many qubits can be read out simultaneously.

So far, the relaxation of the qubits was neglected. To include relaxation, the equations
\begin{equation}
    \dot \sigma_z^{(\mu)} = - (\Gamma_i^{(\mu)} + \Gamma_p^{(\mu)}) \sigma_z^{(\mu)}
\end{equation}
for the qubit populations are taken into account along with Eqs.~(\ref{eq_adot}) and (\ref{eq_sigma-dot}). Here, $\Gamma_i^{(\mu)} = 1/T_1^{(\mu)}$ is the intrinsic relaxation rate of qubit $\mu$ and $\Gamma_p^{(\mu)} = \kappa (g_s^{(\mu)} / \Delta_\mu)^2$ is the Purcell qubit relaxation rate due to the spin-photon coupling (assuming $\bar N \ll 1$)~\cite{PhysRevB.100.245427}. The solution $\taout$ of this extended set of equations has the same form as Eqs.~(\ref{eq_taout}), (\ref{eq_eta}), with an altered $\sigma_z^{(\mu)}$-term in $\eta_\mu(s)$, Eq.~(\ref{eq_eta}), capturing the effect of the relaxation. This term does not depend on the qubit states and thus reduces the readout contrast. We consider the ratio of qubit state-dependent and qubit state-independent terms, and conclude that the effect of relaxation on $\taout(s)$ can be neglected if for all $\mu$
\begin{equation}
    R=\left| \sum_\nu \frac{\sqrt{\kappa_1} \ain g_s^{(\nu)}}{\Delta_\nu s_r^{(\mu)} \left( \frac{1}{2} + \frac{s_r^{(\mu)}}{\Gamma_i^{(\nu)} +\Gamma_p^{(\nu)}} \right)} \right| \ll 1.\label{eq_relaxation}
\end{equation}

For a constant qubit population on the timescale of the readout, $\Gamma_i^{(\mu)}+\Gamma_p^{(\mu)}\to 0$ and $R\rightarrow 0$. With increasing relaxation, the phase shift between the two signals of the output field is reduced and approaches a saturation value. If Eq.~(\ref{eq_relaxation}) is not fulfilled, the effect of the relaxation is too strong and the qubit state-dependent components cannot be identified from the decomposition of the output field. Another limitation for the relaxation rate is $\Gamma_i^{(\mu)}+\Gamma_p^{(\mu)}<1/t_r$, otherwise only limited information is gained. For realistic silicon spin-qubit parameters (\ref{eq_relaxation}) is satisfied and relaxation is expected to be insignificant~\cite{StrongCouplingPrinceton}.

Finally, we compare our proposed transient multi-qubit readout to the conventional dispersive readout of multiple qubits. Conventional dispersive multi-qubit readout has been demonstrated with up to two transmon qubits~\cite{Nature449.443,PhysRevLett.102.200402} but remains elusive for spins so far. Using the dispersive Hamiltonian, Eq.~(\ref{eq_Hdispersive}), conventional multi-qubit readout can be modelled by defining intervals in the range of the observed quadrature associated with each multi-qubit state. The interval the observable is found in will determine which state is assigned to the system. The probability to find the observable in the correct interval after a certain sampling time can be estimated in complete analogy to the discussion with only two intervals described in this section. Comparing to our readout time $t_r$ we find that the transient dispersive readout is approximately one order of magnitude faster if a comparable fidelity $F_m = F_r^{(1)}F_r^{(2)}$ for the multi-qubit state is required. Compared to the readout time of the dispersive readout of a single qubit per resonator we still find a speed-up by a factor of between 1.5 and 2~\cite{PhysRevB.100.245427}. We furthermore emphasize that our proposed transient readout is easier to scale up beyond the two-qubit case since the responses from different are easily separated, whereas conventionally the dispersive shifts must be chosen in a way such that all $\sum_\mu \sigma_\mu \chi_s^{(\mu)}$, $\sigma_\mu = \pm 1$, are unique. Note however, that the transient dispersive readout is either destructive and requires re-initialization or an overhead in qubits and two-qubit gates which limits the readout fidelity.

\section{Conclusions\label{sec_conclusion}}

In summary, we have proposed a protocol for the simultaneous readout of multiple qubits dispersively coupled to a single resonator mode. The contributions of the individual qubits are separated by recording the time-resolved output field and then performing a Laplace transform. The transient dispersive readout proposed in this letter can be a potent tool for the fast, multiplexed, and/or selective readout of resonator-coupled qubits. This holds the promise of greatly advancing the scalability of solid-state qubits since it speeds up the readout of entire registers, leaving more time for gate operation and opening the pathway for quantum error correction. Nonetheless, it is still possible to read out individual qubits if required, by keeping the others detuned from the resonator frequency.

The use of a single resonator is particularly beneficial for semiconductor spin qubits, where it is unrealistic to include a large number of readout resonators into the chip due to their large footprint. While a promising readout fidelity was estimated in this article with realistic parameters for electron spin qubits in Si/SiGe, it can be expected that our proposal will also be feasible for SiMOS~\cite{Guo2023} or hole spin qubits~\cite{Yu23}. We further see potential of our technique in the field of superconducting qubits by reducing the wiring complexity of the readout apparatus.

After demonstrating the basic concept of the readout scheme, we expect that the evaluation protocol can be optimized further for future applications, relying on a fit to the heterodyne signal, rather than the evaluation at certain points in the complex frequency domain.


\begin{acknowledgments}
We thank Jonas Mielke, Joris Kattem\"olle, Adam R. Mills, Xuanzi Zhang, and Jason R. Petta for helpful discussions. This work has been supported by the Army Research Office (ARO) grant number W911NF-15-1-0149.
\end{acknowledgments}


\bibliography{literature.bib}

\end{document}